\documentclass[trackchanges]{aastex6}
\usepackage{xcolor}
\usepackage{textcomp}
\usepackage{hyperref}
\usepackage{natbib} 
\usepackage{color}
\usepackage{multirow}
\usepackage{amsmath}
\usepackage{subfigure}
\usepackage[innercaption]{sidecap}

\newcommand\period{0.94145287}
\newcommand\perioderrorplus{6.56\times10^{-7}}
\newcommand\perioderrorminus{6.59\times10^{-7}}

\newcommand\qlowerlimit{1$\times10^{6}$}
\newcommand\rstara{0.2789}
\newcommand\mpmstar{0.007843}
\newcommand\lschisq{1.07}

\slugcomment{Draft for the Astrophysical Journal}

\shorttitle{Searching For Rapid Orbital Decay of WASP-18b} \shortauthors{Wilkins et al.}

\begin{document}

\title{Searching For Rapid Orbital Decay of WASP-18b}


\author{Ashlee~N.~Wilkins\altaffilmark{1},~Laetitia~Delrez\altaffilmark{2},~Adrian~J.~Barker\altaffilmark{3},~Drake~Deming\altaffilmark{1},\\
~Douglas~Hamilton\altaffilmark{1},~Michael~Gillon\altaffilmark{4},~and~Emmanuel~Jehin\altaffilmark{4}}
\altaffiltext{1}{Department of Astronomy, University of Maryland, College Park, MD 20742-2421, awilkins@astro.umd.edu}
\altaffiltext{2}{Cavendish Laboratory, University of Cambridge, J J Thomson Avenue, Cambridge CB3 0HE}
\altaffiltext{3}{Department of Applied Mathematics, School of Mathematics, University of Leeds, Leeds, LS2 9JT, UK}
\altaffiltext{4}{Space Sciences, Technologies and Astrophysics Research (STAR) Institute, Universit\'{e} de Li\`{e}ge, all\'{e}e du 6 Ao\^{u}t 19C, 4000 Liège, Belgium}


\begin{abstract}
The WASP-18 system, with its massive and extremely close-in planet, WASP-18b (M$_{p}$ = 10.3M$_{J}$, a = 0.02 AU, P = 22.6 hours), is one of the best known exoplanet laboratories to directly measure Q', the modified tidal quality factor and proxy for efficiency of tidal dissipation, of the host star. Previous analysis predicted a rapid orbital decay of the planet toward its host star that should be measurable on the time scale of a few years, if the star is as dissipative as is inferred from the circularization of close-in solar-type binary stars. We have compiled published transit and secondary eclipse timing (as observed by WASP, TRAPPIST, and Spitzer) with more recent unpublished light curves (as observed by TRAPPIST and HST) with coverage spanning nine years. We find no signature of a rapid decay. We conclude that the absence of rapid orbital decay most likely derives from Q' being larger than was inferred from solar-type stars, and find that Q'\,$\geq$\,\qlowerlimit, at 95\,\% confidence; this supports previous work suggesting that F-stars, with their convective cores and thin convective envelopes, are significantly less tidally dissipative than solar-type stars, with radiative cores and large convective envelopes.
\end{abstract}


\section{Introduction}

The discovery of WASP-18b \citep{Hellier2009,Southworth2009}, with its large mass (10.3\,M$_{J}$) and small orbit (22.6\,hours) (see Table~\ref{tab:system_details} for other parameters), elicited one primary question: how could it exist? A planet of that mass and proximity should raise a substantial tidal distortion (tidal bulge) in the central star. Because the star is not a perfectly elastic body, and because the planet orbits more quickly than the star rotates, the tidal bulge would lag behind the planet, causing the planet's orbital motion to accelerate and the orbit to shrink \citep{Goldreich1966}. \citet{Hellier2009} calculated a 0.65\,Myr future lifetime for the planet, assuming that the star is as dissipative as is inferred from the circularization of close solar-type binary stars \citep{Meibom2005,Ogilvie2007}. The estimated age of the star is a few hundred million years \citep{Bonifanti2016,Hellier2009} to 2 Gyr \citep{Southworth2009}; finding a planet with a lifetime that is such a small fraction of the system's age is extremely improbable. \citet{Hamilton2009} discusses several alternative explanations, ranging from an overestimation of the decay rate (due to unmodeled nuances of tidal physics leading to an underestimation of the tidal Q' parameter) to a non-tidal mechanism holding the planet in place (e.g. influence of another body in the system). \citet{Barker2009} investigate the efficiency of tidal dissipation in the convective envelopes of F-stars, which have both convective cores and convective envelopes; G stars, on which most studies of exoplanetary tidal decay focus (e.g. \citealt{Jackson2009,Birkby2014}), have radiative cores and thicker convective envelopes. The \citet{Barker2009} calculations reveal that tidal dissipation within F stars is generally much less efficient than within G-stars, and therefore that planetary tidal decay around stars like WASP-18 would be imperceptibly low over a decadel timespan \citep{Barker2010,Barker2011a}. If, however, tidal dissipation within WASP-18 behaved as is usually inferred for solar-type stars, \citet{Birkby2014} predict that its transit should occur progressively earlier at each observation, accumulating to a measurable shift of nearly six minutes over ten years. This is the largest predicted shift of any planet, making the WASP-18 system possibly the best known laboratory for direct measurements of the stellar tidal Q' parameter. \citet{Maciejewski2016} potentially measured the tidal decay of WASP-12b, but \citet{Hoyer2016b} ruled out the orbital decay of WASP-43b proposed by \citet{Jiang16}. 

In this Letter, we bring together published measurements of transit and secondary eclipse timing from discovery \citep{Hellier2009}, \emph{Spitzer} \citep{Nymeyer2011,Maxted2013}, and ground-based TRAPPIST \citep{Maxted2013} observations, and new analyses of unpublished archival (HST), and recent TRAPPIST data. We place strong limits on the maximum rate of the system's orbital decay and discuss the implications.


\section{New Observations}\label{sec:observations} 

\subsection{TRAPPIST}

The TRAnsiting Planets and PlanetesImals Small Telescope - South (TRAPPIST-S, \citealt{Jehin2011,Gillon2011}) is a ground-based, 60-cm robotic telescope based at the La Silla Observatory used to study both exoplanets and small bodies in the Solar System. TRAPPIST observed two WASP-18b photometric transits in the fall of 2015 in the broad-band Sloan-z filter, centered at 0.9134\,$\mu$m. 

\subsection{\emph{Hubble Space Telescope}}
The \emph{Hubble Space Telescope} (HST) observed WASP-18b in 2014 in spatial scan mode \citep{Deming2013} over its full phase (PID 13467, PI Bean), including one full transit, one full secondary eclipse, and one extra eclipse ingress. All observations were made with the Wide Field Camera 3 (WFC3) G141 infrared grism, covering 1.1 - 1.7\,$\mu$m. While the primary deliverable from such observations is the spectrum, we sum over wavelength to extract a photometric light curve. To maximize observing efficiency, the scan reverses direction, rather than taking the time to reset to the starting point, at the end of each scan. This introduces a non-constant offset requiring separate analysis of the forward and reverse scans.  


\section{Analysis: Deriving the new White Light Curves}

Table~\ref{tab:all_obs} includes the transit and secondary eclipse times used in this analysis. We describe here how we generated white light curves and transit fits to the new TRAPPIST and HST data.

\subsection{TRAPPIST Light Curves}
\begin{figure*}
\begin{center}
    \begin{subfigure}
        \centering
        \includegraphics[height=4.2in]{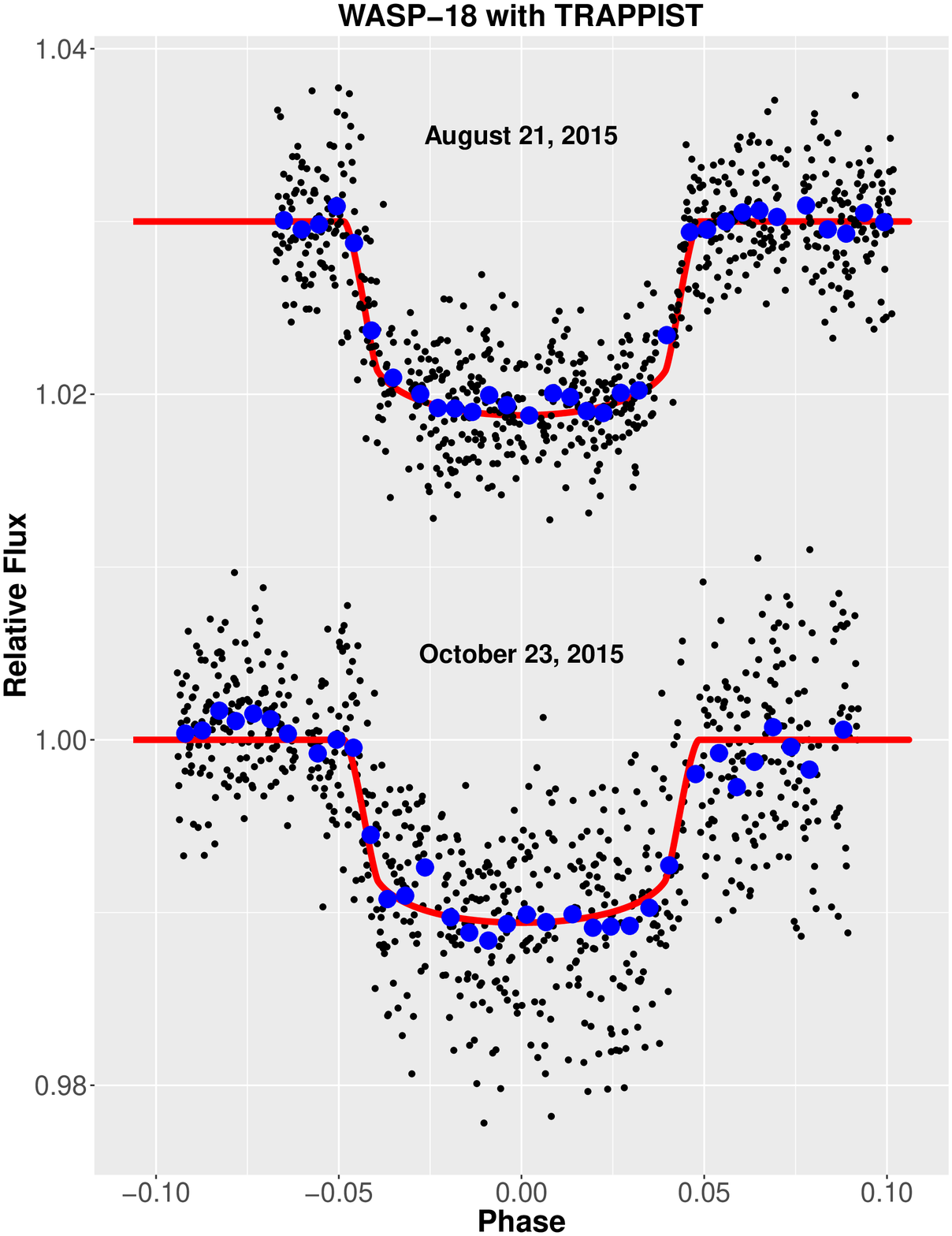}
    \end{subfigure}
    ~ 
    \begin{subfigure}
        \centering
        \includegraphics[height=4.2in]{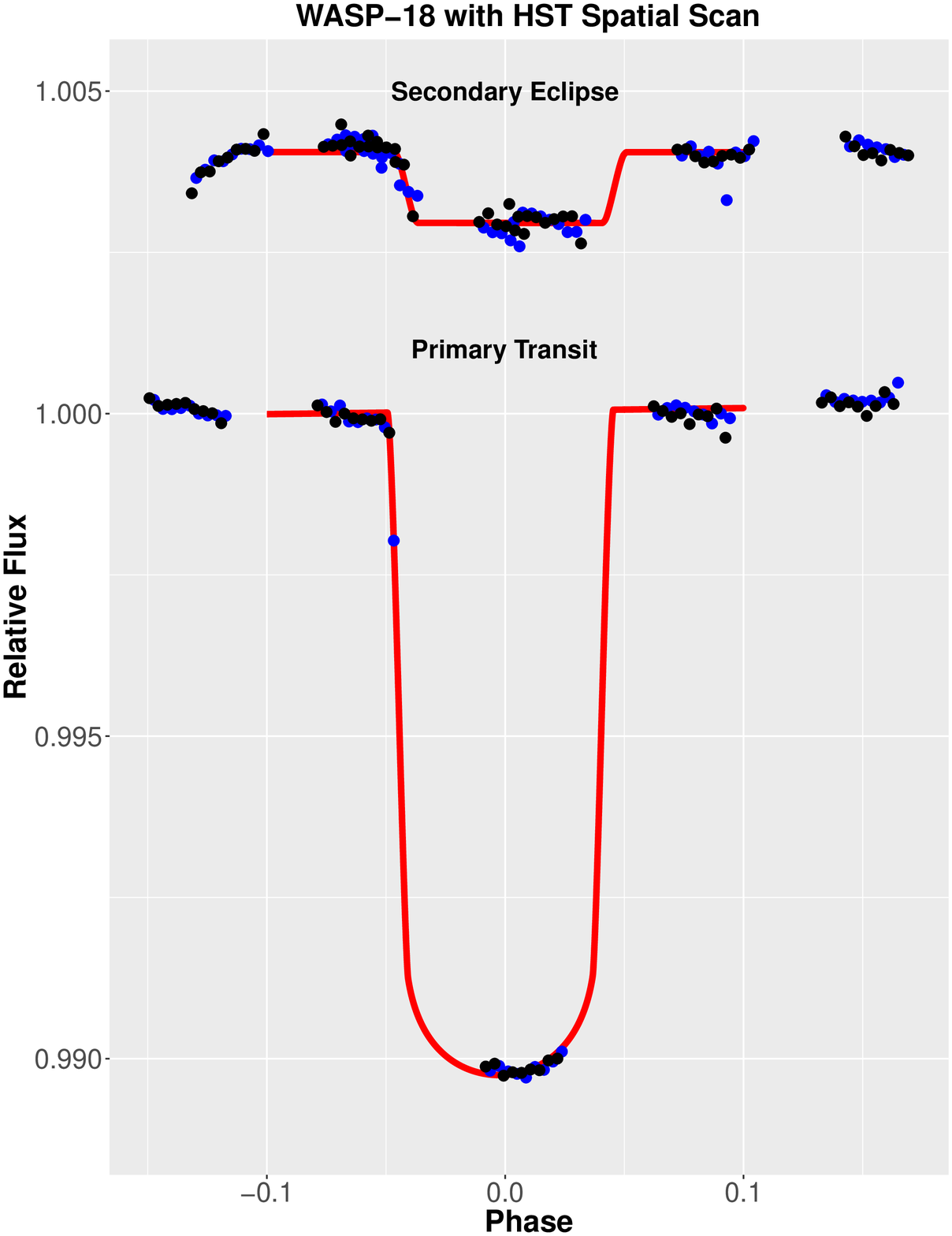}
    \end{subfigure}
    \caption{\emph{Left:} Two new transits of WASP-18 by its planet observed by TRAPPIST in 2015. The data are plotted in black, binned points in blue, and the best-fit transit curve in red. The August observation is offset in y for visualization purposes. \emph{Right}: New transit and secondary eclipse of WASP-18 by its planet observed by HST in 2014. The data are plotted in black and blue (forward and reverse scans), the best-fit transit curve in red.}
    \label{fig:whitelights}
\end{center}
\end{figure*}

We reduce our TRAPPIST data in the methods described by \citet{Gillon2012}. We calculated the best-fit transit curve for each observation using the TRAPPIST MCMC procedure (\citealt{Gillon2009} and references therein), executing the \citet{Mandel2002} algorithm to find the new best-fit light curve parameters. We generated the curve plotted in Figure~\ref{fig:whitelights} with the BATMAN procedure \citep{Kreidberg2015}, given those orbital parameters. 

\subsection{HST White Light Curves}
As has been studied extensively (e.g. \citealt{Sing2016}), the HST WFC3 camera, while improved over its predecessor NICMOS, has persistent systematic errors that seem to be a function of incident flux \citep{Wilkins2014}, with three distinctive effects: a visit-long ramp, an orbit-long ramp, and a ``hook" within orbits \citep{Berta2012,Wilkins2014}. We reduce the WFC3 data and mitigate systematics in a modified divide-oot method -- a method of averaging out all three systematic effects \citep{Wilkins2014,Deming2013}, including the correction to the STScI wavelength calibrations found in \citet{Wilkins2014}. To fit the transit, we use the non-linear, fourth-order limb darkening coefficients from \citet{Claret2000} in the \citet{Mandel2002} light curve models, and derive ``prayer-bead" error bars as in \citet{Gillon2009}. To fit the the secondary eclipse, we use the same procedure in the limit of no limb darkening, such that the shape is that of a trapezoid. We analyze the forward and reverse scans independently, as mentioned in \S~\ref{sec:observations}, due to a non-linear offset between the two; the final timing results agree and are thus shown as an average in the table. 

\section{Results: Transit Timing Evolution over Nine Years}\label{sec:results}

We have compiled all published transit and secondary eclipse observations of WASP-18b and added them to the new observations obtained by  HST and TRAPPIST to produce a data set spanning more than nine years. The full data set is found in Table~\ref{tab:all_obs}. Published light curve solutions came from four observing campaigns: 

\emph{WASP:} The Wide-Angle Search for Planets (WASP) Project \citep{Pollacco2006} announced the discovery and initial orbital solution of WASP-18b as observed in transit by the WASP-South Survey and in radial velocity with the CORALIE spectrograph \citep{Hellier2009}, and confirmed with the Danish 1.5m telescope at ESO \citep{Southworth2009}. The \citet{Southworth2009} ephemeris was later found to be erroneous \citep{Southworth2010}; we use only the \citet{Hellier2009} ephemeris.

\emph{Spitzer:} \citet{Nymeyer2011} observed two secondary eclipses of WASP-18b via the \emph{Spitzer} Exoplanet Target of Opportunity Program with the Infrared Array Camera (IRAC, PID 50517). The first secondary eclipse was observed in the 3.6\,$\mu$m and 5.8\,$\mu$m channels on December 20, 2008, the second in the 4.5\,$\mu$m and 8.0\,$\mu$m channels on December 24, 2008. \citet{Maxted2013} reanalyzed the \citet{Nymeyer2011} points. 

\emph{Warm \emph{Spitzer}:} \citet{Maxted2013} observed two full phases of WASP-18b's orbit with warm \emph{Spitzer}, one with the 3.6\,$\mu$m channel on January 23, 2010, and the other with the 4.5\,$\mu$m channel on August 23, 2010.

\emph{TRAPPIST:} In addition to the unpublished, new transit curves presented as part of this work, TRAPPIST also observed WASP-18b five times in transit in late 2010 and early 2011, also in the Sloan-z' filter \citet{Maxted2013}. 

\begin{figure*}
\begin{center}
\includegraphics[width=4.5in]{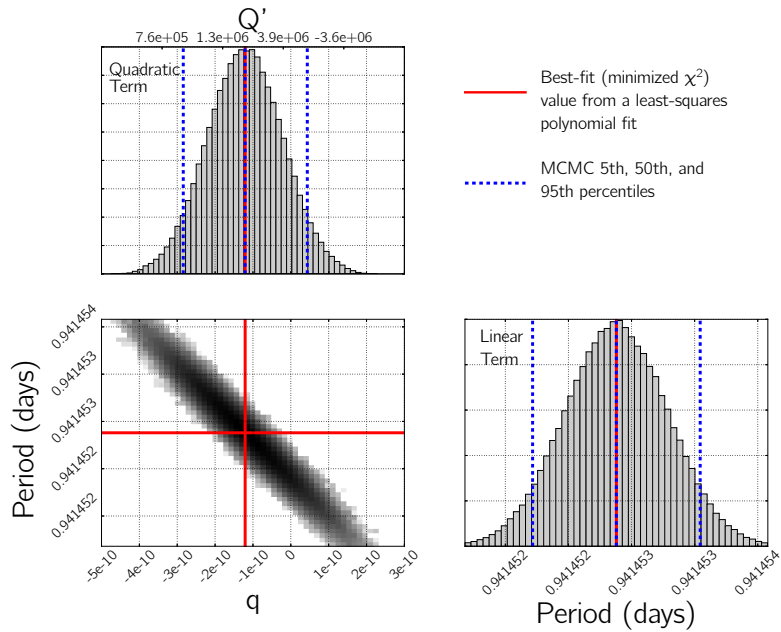}
\caption{MCMC posterior probability distributions for the linear and quadratic parameters of the quadratic fit, $q$ (proportional to -1/Q'), and $p$ (orbital period), with 5th, 50th, and 95th percentiles marked by the dashed lines. We leave the less important intercept term off of this corner plot, for clarity. Overplotted in red are the best-fit values from the least-squares polynomial fit (minimizing $\chi^{2}$). The two methods agree on the value of the period and they both find only an upper limit for the magnitude of the quadratic term (corresponding to a lower limit on Q', see top axis of the $q$ plot.}
\label{fig:corner}
\end{center}
\end{figure*}

To search for tidal decay, we study the correlation between the number of orbits since discovery ephemeris and transit (or eclipse) arrival time. In the case of no orbital evolution, this correlation would be linear, and the slope of the line would be the planetary orbital period. We allow for the possibility of decay by including a second-order term that is dependent on the rate of any orbital evolution. We first perform a multivariate linear regression and find a plausible fit ($\chi^{2}_{RED} = \lschisq$). To explore the trade-off between the linear and quadratic terms of the fits, we also perform a Markov Chain Monte Carlo (MCMC) quadratic fit using $emcee$ \citep{Foreman-Mackey2013}; the results of both fits, which are in excellent agreement, are in Figure~\ref{fig:corner}. With $emcee$, we find the period P = $\period^{+\perioderrorplus}_{-\perioderrorminus}$\,days, in agreement with the \citealt{Hellier2009} P = 0.94145299\,$\pm$\,8.7$\times10^{-7}$\,days. If WASP-18 were as tidally dissipative as is inferred from the circularization of solar-type close binary stars, there should be a definitive deviation from linear behavior, i.e., the quadratic term should be nonzero. We measure an upper limit for the magnitude of the quadratic term, and we therefore find no confirmation of rapid tidal decay for the WASP-18 system. Indeed, as discussed in the next section, we should not have expected to find evidence of rapid decay.

\section{Discussion: Implications of the Absence of Rapid Tidal Decay}\label{sec:discussion}

\edit3{Without strong evidence of a rapidly decaying orbit} suggested by \citet{Hellier2009,Birkby2014}, we turn instead to the predictions of \citet{Barker2009,Barker2010,Barker2011a,Lanza2011}. We first briefly review the discussion of these predictions as they apply to WASP-18, and then calculate a constraint on the Q' of WASP-18. 

\subsection{Tidal Dissipation in G vs. F Stars}
Tides raised within a central star by a planetary companion are dissipated within the star and angular momentum is transferred between the stellar spin and planetary orbit in the process (e.g. \citealt{Ogilvie2014}). For short-period planets (orbiting sub-synchronously rotating stars), like WASP-18b, that also have approximately circular orbits, tidal dissipation in the star causes the planet to lose angular momentum and spiral inward, because the tidal bulge raised in the star lags the planet when the planet's orbital period is less than the star's rotational period (i.e. P$_{orb}$ \textless~P$_{rot}$). This is the opposite of the Earth-Moon system, in which the Moon recedes from the Earth because the bulge leads the Moon (since P$_{orb}$ \textgreater P$_{rot}$). The rate of change of the orbit depends on the efficiency of tidal dissipation within the host star; this is where stellar structure becomes important. 

The tide in the star is often decomposed into two contributions: an equilibrium tide and a dynamical tide (e.g. \citealt{Zahn1977}). Dissipation of both components is expected to become less efficient in stars slightly more massive than the Sun (i.e. F stars). While we often generalize Sun-like stars (typically defined as 0.5M$_{\odot}$ $\lesssim$ M$_{*}$ $\lesssim$ 1.3 M$_{\odot}$) to have radiative cores and convective envelopes and more massive stars to have the opposite, development of convective cores and radiative envelopes is actually a continuum. WASP-18, for example, is a 1.2-M$_{\odot}$ F6 star, and, according to MESA stellar structure models \citep{Paxton2011}, should have a convective core within the innermost 6\,\% of the stellar radius, and a convective envelope in the outer 15\,\%; it is therefore intermediate between an solar-mass and high-mass star. For tidal dissipation, therefore, an F star like WASP-18 is not "sun-like." 

We quantify the efficiency of tidal dissipation using the tidal quality factor, Q, defined as \citep{Goldreich1963}:
\begin{equation}
    Q \equiv \frac{\text{energy stored in tidal distortion}}{\text{energy dissipated in one cycle}} = 2\pi E_{0}(\oint-\dot{E} dt)^{-1},
\end{equation}
where $E_{0}$ is the maximum energy stored in the tidal bulge and $\dot{E}$, intrinsically negative, is the energy dissipated in one tidal period. We use the modified Q (i.e. Q') convention throughout this Letter:
\begin{equation}
Q_{*,0}^{'} \equiv \frac{3Q}{2k_{2}},
\end{equation}
where $k_{2}$ is the tidal Love number. Q' is almost certainly not a single constant number for all stars (even of the same spectral type), but is instead a complicated function of the stellar mass, structure, rotation, and tidal periods, as well as the planetary properties (e.g. \citealt{Ogilvie2014}). Q' is the Q of an equivalent homogeneous body ($k_{2}$\,=\,3/2). A large Q' corresponds to weak or inefficient tidal dissipation, and a smaller Q' corresponds to strong or efficient dissipation. We investigate here the Q' of the star (WASP-18), not the planet (WASP-18b); the planet's tidal Q' is relevant for its own tidal evolution, and leads to synchronization of its rotation and circularization of its orbit. 

The equilibrium tide is dissipated within the convective envelope of the star by the effective viscosity of convective turbulence \citep{Zahn1966,Goldreich1977}; however, the effective viscosity may be significantly reduced in the case of a short-period planet \citep{Penev2011,Ogilvie2012}. In addition, in F stars, the outer convection zone is thin and of very low mass, so it is expected to be much less dissipative than in G-stars; the effective tidal Q' could be as high as 10$^{11}$ \citep{Barker2009} for a star like WASP-18 at the tidal frequencies of interest. Dissipation in the convective core of an F star is also likely weak (e.g. \citealt{Zahn1977}). 

The dynamical tide primarily consists of internal gravity (g-mode) waves that are tidally excited at the convective-envelope-radiative-core boundary and propagate inwards to the center of the star. These waves are thought to be damped by radiative diffusion or nonlinear effects. If they can reach the center, they become geometrically focused and, if the planet exciting them is sufficiently massive -- like WASP-18b -- they may reach sufficiently large amplitudes such that they break, leading to significantly enhanced tidal dissipation \citep{Goodman1998,Ogilvie2007,Barker2010,Barker2011a}. This process deposits angular momentum into the star, thereby removing angular momentum from the planet's orbit; the star's rotation gets faster (``spin-up"), while the planet's orbit shrinks. If WASP-18 were Sun-like, WASP-18b would be sufficiently massive to cause wave breaking, and we would expect the planet to rapidly spiral into its star. However, in the case of an F-star like WASP-18, the convective core prevents the tidally-excited gravity waves from reaching the center where they would be focused, so that they may never reach such large amplitudes to break, though they may be subject to weaker nonlinear effects (eg. \citealt{Barker2011b,Weinberg2012,Essick2016}). The dissipation would be significantly reduced, save for select resonant tidal frequencies, so that we would expect the planet to remain in the orbit in which it was discovered \citep{Barker2009}. Furthermore, the lingering thin outer convective envelope in an F star of WASP-18's mass would inhibit radiative damping of the waves near the top of the radiative zone (relative to more massive A-stars). The dissipation that \citet{Valsecchi2014} find for WASP-71 may be moderately higher than we would expect in WASP-18 precisely because it is a more massive (1.5 versus 1.2\,$M_{\odot}$) star, and therefore has a thinner outer convective envelope than WASP-18, but what they obtain is still very weak. 

 Were a resonance present, Q' could indeed be very low, and therefore the star could be quite dissipative. However, the above arguments and those of, e.g. \citet{Lanza2011,Barker2009}, support a high-Q', generally minimally-dissipative scenario for a star like WASP-18. 

\subsection{Estimating the Tidal Q' for WASP-18}\label{sec:tidalq}

When a planet transits its host star, we have a convenient time point from which to measure any changes in the orbit, which we infer through a shift in the transit (or secondary eclipse) arrival time. \citet{Birkby2014} show that the expected shift can be reduced to: 
\begin{equation}
    T_{shift} = -\left(\frac{27}{8}\right)\left(\frac{M_{p}}{M_{*}}\right)\left(\frac{R_{*}}{a}\right)^{5}\left(\frac{2\pi}{P}\right)\left(\frac{1}{Q_{*}{'}}\right)T^{2},
    \label{tshift}
\end{equation}
where $M_{p}/M_{*}$ is the planet-to-star mass ratio (for WASP-18, $M_{p}/M_{*}$\,=\,\mpmstar), $R_{*}/a$ is the stellar-radius-to-semi-major-axis ratio (for WASP-18, $R_{*}/a$\,=\,\rstara), T is the elapsed time, and P is the orbital period of the planet. Therefore, in a quadratic fit of the form
\begin{equation} 
t = qT^{2} + pT + c,
\label{eq:quadratic}
\end{equation}
where the linear coefficient $p$ corresponds to the period of the planet's orbit, the quadratic term is defined by Equation~\ref{tshift}. Rearranging, we find that Q' depends on the quadratic coefficient $q$ as: 
\begin{equation}
    Q' = -\left(\frac{27}{8}\right)\left(\frac{M_{p}}{M_{*}}\right)\left(\frac{R_{*}}{a}\right)^{5}\left(\frac{2\pi}{P}\right)\left(\frac{1}{q}\right)
\end{equation}
We fit for the coefficients in Equation~\ref{eq:quadratic}, and thus the period and Q', as discussed in \S~\ref{sec:results} and shown in Figure~\ref{fig:corner}.

Equation~\ref{tshift} makes clear that a planet must be close-in and massive (relative to the radius and mass of the host star), its orbital period must be short, and it must orbit a star with a favorable $Q{'}$, in order to produce any discernible shift in time. Currently, in the NASA Exoplanet Archive, only eight confirmed planets have both masses larger than 1.0\,$M_{J}$ and orbital periods of roughly one day or less. The addition of recently-announced KELT-16b \citep{Oberst2016} makes nine. Of those, one is around a pulsar and four (WASP-18b, KELT-16b, WASP-12b, and WASP-103b) are around stars more massive than 1.2\,$M_{\odot}$ and therefore likely possessing convective cores that preclude tidal wave breaking at the center. Of the remaining four, WASP-43b has already demonstrated no rapid tidal decay \citep{Hoyer2016b}, but WASP-19, WTS-2, and K2-22 all orbit stars less massive than the sun, and may be reasonable testbeds for dissipation within a star with a larger convective envelope and a smaller radiative core; \citet{Birkby2014} has already suggested that WTS-2 should have a barely-discernible shift for Q'=10$^{6}$ (17 s over 16 years). 

Equation~\ref{tshift} assumes a stellar obliquity of zero and neglects tidal dissipation in the planet, assuming its orbit to be circularized and its spin to be synchronized and aligned with the orbit. The canonical value of Q' is 10$^{6}$, as derived for stars from measurements of the orbits of binary star systems (e.g. \citealt{Meibom2005}) and for Solar System giant planets from the orbits of their satellites \citep{Zhang2008}). 

We return to Figure~\ref{fig:corner}, as we can now interpret the findings for $q$ (and therefore Q') physically. The 95th percentile posterior probability distribution for q is effectively zero; given that Q'\,$\propto$\,$\frac{1}{q}$, this means we only can definitively extract a lower limit, Q'$\geq$\qlowerlimit, taken at the 5th percentile of the $q$ distribution. Continued monitoring of this system should further constrain WASP-18's Q', and it follows from the discussion above that we will continue to find an increasing lower limit, i.e. no evidence of rapid tidal decay.


\section{Conclusion} 
We have combined previously published and new data to find no conclusive evidence of rapid tidal decay of the orbit of WASP-18b, supporting predictions of little to no tidal decay for a short-period planet around an F star \citep{Barker2009,Barker2010,Barker2011a}, given our current understanding of the physics of tidal dissipation in F stars. We find for WASP-18 that Q'\,$\geq$\,\qlowerlimit at 95\,\% confidence. Further observations of WASP-18b and similar monitoring of planets like WASP-19b, WTS-2b, and K2-22b would add tighter observational constraints on stellar Q' for various stellar types, and allow us to further probe the mechanisms of stellar tidal dissipation. 


\section*{Acknowledgements}
The authors thank the referee for a careful eye and helpful insights. TRAPPIST is a project funded by the Belgian Fund for Scientific Research (F.R.S.-FNRS) under grant FRFC 2.5.594.09.F, with the participation of the Swiss National Science Foundation (SNF). L.D. acknowledges support of the F.R.I.A. fund of the F.R.S-FNRS. M. Gillon and E. Jehin are F.R.S.-FNRS Research Associates. AJB was supported by The Leverhulme Trust through the award of an Early Career Fellowship. 


\clearpage

 \begin{deluxetable*}{cccc}

    \tablecaption{WASP-18 parameters used for this analysis.}
    \tablehead{\colhead{\bf{Parameter}} & \colhead{\bf{Value}} & \colhead{\bf{Average}} & \colhead{\bf{Reference}}}
    \startdata
        \multicolumn{4}{c}{The Star: WASP-18}\\
        \vspace{-0.15cm} & & & \\ 
        \hline
        \vspace{-0.15cm} & & & \\ 
        \vspace{-0.15cm} & 1.29\,$\pm$\,0.16 & & \citealt{Doyle2013} \\ 
        \vspace{-0.15cm}
        Radius (R$_{\odot}$) & & 1.22\,$\pm$\,0.11 & \\
         & 1.15\,$\pm$\,0.02 & &  \citealt{Bonifanti2016} \\
        \vspace{-0.15cm} & & & \\ 
        \hline
        \vspace{-0.15cm} & & & \\ 
        & 6400\,$\pm$\,100 & &  \citealt{Hellier2009} \\
        T$_{eff}$ (K) & 6400\,$\pm$\,75 & 6322\,$\pm$\,72 & \citealt{Doyle2013} \\
        & 6167\,$\pm$\,7 & & \citealt{Bonifanti2016} \\
        \vspace{-0.15cm} & & & \\ 
        \hline
        \vspace{-0.15cm} & & & \\ 
        & 4.4\,$\pm$\,0.15 & & \citealt{Hellier2009} \\
        log\,$g$\, & 4.32\,$\pm$\,0.09 & 4.32\,$\pm$\,0.10 & \citealt{Doyle2013} \\
        & 4.39\,$\pm$\,0.01 & &  \citealt{Bonifanti2016} \\
        \vspace{-0.15cm} & & & \\ 
        \hline
        \vspace{-0.15cm} & & & \\ 
        & \textless\,2.0 & & \citealt{Southworth2009} \\
        Age (Gyr) & 0.5 - 1.5 & & \citealt{Hellier2009} \\
        & 0.9\,$\pm$\,0.2 & & \citealt{Bonifanti2016}\\
        \vspace{-0.15cm} & & & \\ 
        \hline
        \vspace{-0.15cm} & & & \\ 
        \vspace{-0.15cm} & 1.281$^{+.052}_{-.046}$ & & \citealt{Southworth2009} \\
        M$_{*}$ (M$_{\odot}$) & 1.24\,$\pm$\,0.04 & 1.25\,$\pm$\,0.04 & \citealt{Triaud2010} \\
        \vspace{-0.15cm} & 1.22\,$\pm$\,0.03 & & \citealt{Enoch2010} \\
        \vspace{-0.15cm} & & &  \\ 
        \hline
        \vspace{-0.15cm} & & &  \\
         \multicolumn{4}{c}{The Planet: WASP-18b}\\
        \vspace{-0.15cm} & & & \\ 
        \hline
        \vspace{-0.15cm} & & & \\ 
        P (days) & 0.94145299 \,$\pm$\,8.7$\times10^{-7}$ & & \citealt{Hellier2009}\\ 
        \vspace{-0.15cm} & & &  \\ 
        \hline
        \vspace{-0.15cm}& 0.02047$^{+.00028}_{-.00025}$ & & \citealt{Southworth2009} \\
        a (AU) & & 0.02034$^{+.00026}_{-.00023}$ &  \\
        \vspace{-0.15cm} & 0.02020$^{+.00024}_{-.00021}$ & & \citealt{Triaud2010}\\
        \vspace{-0.15cm} & & & \\ 
        \hline
        \vspace{-0.15cm} & 86\,$\pm$\,2.5 & & \citealt{Hellier2009} \\ 
        \vspace{-0.15cm}
        $i$ ($^{\circ}$) & & 83.3$^{+1.9}_{-2.0}$ & \\
        \vspace{-0.15cm} & 80.6$^{+1.1}_{-1.3}$ & & \citealt{Triaud2010} \\
        \vspace{-0.15cm} & & & \\ 
        \hline
        \vspace{-0.15cm} & & & \\ 
        \vspace{-0.15cm}& 10.43$^{+.30}_{-.24}$ & & \citealt{Southworth2009} \\
        M$_{p}$ (M$_{J}$) & & 10.27$^{+.27}_{-.23}$ &  \\
        \vspace{-0.15cm} & 10.11$^{+.24}_{-.21}$ & & \citealt{Triaud2010}\\
        \vspace{-0.15cm} & & &  \\ 
        \hline
        \vspace{-0.15cm} & & & \\
        \label{tab:system_details}
    \enddata
\end{deluxetable*}

\begin{deluxetable*}{lllll}
\centering 
\tablecaption{WASP-18 Full Observation Summary} 
\tablehead{\colhead{Facility} & \colhead{Date} & \colhead{Original Reference(s)} & \colhead{Orbit} & \colhead{BJD (TDB)}}
\startdata 
WASP & May - December 2006 & \citet{Hellier2009} & 0 & 2454221.48163$\pm$0.00038\\
Spitzer & December 20, 2008 & \citet{Nymeyer2011,Maxted2013}& 636.5 & 2454820.7168$\pm$0.0007 \\
Spitzer & December 24, 2008 & \citet{Nymeyer2011,Maxted2013} & 640.5 & 2454824.4815$\pm$0.0006\\
Warm \emph{Spitzer} & January 23, 2010 & \citet{Maxted2013} & 1061.5 & 2455220.8337$\pm$0.0006\\
Warm \emph{Spitzer} & January 24, 2010 & \citet{Maxted2013} & 1062 & 2455221.3042$\pm$0.0001 \\
Warm \emph{Spitzer} & August 23, 2010 & \citet{Maxted2013} & 1285.5 & 2455431.7191$\pm$0.0003 \\
Warm \emph{Spitzer} & August 24, 2010 & \citet{Maxted2013} & 1286 & 2455432.1897$\pm$0.0001 \\
TRAPPIST & September 30, 2010 & \citet{Maxted2013} & 1327 & 2455470.7885$\pm$0.00040\\
TRAPPIST & October 2, 2010 & \citet{Maxted2013} & 1330 & 2455473.6144$\pm$0.00090\\
TRAPPIST & December 23, 2010 & \citet{Maxted2013} & 1416 & 2455554.5786$\pm$0.00050\\
TRAPPIST & January 8, 2011 & \citet{Maxted2013} & 1433 & 2455570.5840$^{+0.00045}_{-0.00048}$\\
TRAPPIST & November 11, 2011 & \citet{Maxted2013} & 1758 & 2455876.5559$\pm$0.0013\\
HST & April 22, 2014 & This work & 2840.5 & 2456895.6773$\pm$0.0006\\
HST & April 22, 2014 & This work & 2841 & 2456896.1478$\pm$0.0008\\
TRAPPIST & August 20, 2015 & This work & 3223 & 2457255.7832$^{+0.00030}_{-0.00029}$ \\
TRAPPIST & October 21, 2015 & This work & 3291 & 2457319.8010$^{+0.00039}_{-0.00038}$ \\
\label{tab:all_obs} 
\enddata 
\tablecomments{Transit (whole number cycles) and secondary eclipse (half-integer cycles) central times used and/or calculated in this work. Orbit corresponds to number of orbits since the discovery ephemeris, and BJD (TDB) is the best-fit time for the center of the transit or secondary eclipse.} 

\end{deluxetable*} 

\end{document}